\documentclass[aps,pra,reprint,floatfix]{revtex4-1}
\usepackage[utf8]{inputenc}
\usepackage{amsfonts}
\usepackage{amssymb}
\usepackage{amsmath}
\usepackage{amstext}
\usepackage{graphicx}
\usepackage{wasysym}
\usepackage{hyperref}

\newcommand{\Ie}{\textit{i.e.}~}

\newcommand{\I}{{\rm i}}

\renewcommand{\d}{{\rm d}}
\newcommand{\h}{\hslash}
\newcommand{\kb}{k_{\rm B}}

\newcommand{\x}{i}
\newcommand{\y}{j}
\newcommand{\z}{\sigma}

\renewcommand{\o}[2][]{\hat{#2} ^{\vphantom{\dagger}}_{#1}}
\newcommand{\op}[2][]{\hat{#2} ^{\dagger}_{#1}}
\newcommand{\oo}[2][]{\hat{#2} ^{\vphantom{\dagger}2}_{#1}}
\newcommand{\oop}[2][]{\hat{#2} ^{\dagger 2}_{#1}}

\begin{document}

\title{An ultracold analogue to star formation: Spontaneous concentration of energy in trapped quantum gases}

\author{M.\,P.\,Strzys}
\email{strzys@physik.uni-kl.de}
\author{J.\,R.\,Anglin}
\affiliation{OPTIMAS Research Center and Fachbereich Physik, Technische Universit\"at Kaiserslautern, D--67653 Kaiserslautern, Germany}


\begin{abstract}
Stars form when cold cosmic nebulae spontaneously develop hot spots that steadily intensify until they reach fusion temperatures \cite{Schw58}. Without this process, the universe would be dark and dead. Yet the spontaneous concentration of heat is exactly what the Second Law of Thermodynamics is in most cases supposed to forbid. The formation of protostars has been much discussed \cite{Eddi26,Thir70,Lynd77,Lynd99,Posc05}, for its consistency with the Second Law depends on a thermodynamical property that is common in systems whose strongest force is their own gravity, but otherwise very rare: negative specific heat. Negative specific heat turns the world upside down, thermodynamically; it implies that entropy increases when energy flows from lower to higher energy subsystems, opposite to the usual direction. Recent experiments have reported negative specific heat in melting atomic clusters \cite{Schm01, Gobe02} and fragmenting nuclei \cite{Dago00a}, but these arguably represent transient phenomena outside the proper scope of thermodynamics.  Here we show that the counter-intuitive thermodynamics of spontaneous energy concentration can be studied experimentally with trapped quantum gases, by using optical lattice potentials \cite{Jaks05,Bloc08} to realize weakly coupled arrays of simple dynamical subsystems that share the peculiar property of self-gravitating protostars, of having negative micro-canonical specific heat. Numerical solution of real-time evolution equations confirms the spontaneous concentration of energy in such arrays, with initially dispersed energy condensing quickly into dense `droplets'. We therefore propose laboratory studies of negative specific heat as an elusive but fundamentally important aspect of thermodynamics, which may shed fresh light on the general problem of how thermodynamics emerges from mechanics.
\end{abstract}

\maketitle

The Second Law of Thermodynamics says that entropy cannot decrease, which is often summarized by saying that heat cannot flow from colder systems to hotter. This summary is exact if hotter and colder are interpreted as higher and lower temperature, but it is not necessarily valid in reference to systems containing more or less energy. The rate at which temperature changes with energy --- the \textit{specific heat} $C_{V}$ --- is normally positive, so that higher temperature implies higher energy, and heat flow from high to low temperature redistributes energy more evenly throughout aggregate systems, tending towards uniform equilibrium.  If $C_{V}$ should be negative, however, then higher temperature corresponds to \textit{lower} energy, and increasing entropy requires that systems with less energy lose heat to systems with more energy. In aggregates of many subsystems, negative $C_{V}$ thus implies instability toward spontaneous energy concentration. This would fit the pattern seen in star formation, but that does not prove that the thermodynamic explanation of star formation can only be negative $C_{V}$. It is debated whether or under which circumstances negative $C_{V}$ is really possible \cite{Mich07,Lynd08,Cari10,Rami08,Thir03}.

Some classic thermodynamics texts state flatly that systems with negative $C_{V}$ cannot exist \cite{Pipp57}, while others discuss the issue at more cautious length \cite{Land80}. In fact thermodynamics does not prescribe such features of any system. It is the discipline of statistical mechanics that strives to predict them, by representing the effect of complex mechanical interactions in terms of probability distributions, and identifying statistical properties of these distributions with thermodynamical quantities like $C_{V}$. According to the workhorse probability distribution of statistical mechanics, the so-called canonical ensemble (CE), negative $C_{V}$ is indeed impossible for any system:
\begin{align}\label{CVcanon}
C_V^{\rm CE} = \frac{\d}{\d T}\langle E\rangle_{\rm CE} = \frac{\langle (E - \langle E\rangle)^2\rangle_{\rm CE}}{\kb T^2} > 0,
\end{align}
where the averages $\langle \,\cdot\, \rangle_{\rm CE}$ are taken in the canonical ensemble (see Supplementary Material). This theorem does not hold, however, in the alternative probability distribution of the micro-canonical ensemble ($\mu$C), in which the system's energy is fixed, rather than fluctuating randomly as in the CE. It has been argued that the canonical ensemble is invalid whenever the micro-canonical ensemble yields negative $C_{V}$ \cite{Hert71}.  These discrepancies between ensembles point to the fact that statistical mechanics remains a comparatively weak link in the chain of physics. 

It is nonetheless a tremendously important link. Statistical mechanics gives physics much of its power, by connecting simple models to complex reality, and its predictions are amply verified in many cases. They are based on falsifiable assumptions, however, and these can be tested today as never before, with experiments on tightly controlled mesoscopic systems whose dynamics can be followed closely enough for direct comparison with the statistical theory. Trapped quantum gases offer an especially promising laboratory for such tests. The observation of Bose-Einstein condensation was a scientific landmark because it confirmed one dramatic prediction of statistical mechanics, the cessation of thermal motion by a large fraction of gas atoms at a temperature above absolute zero. Can trapped quantum gases test another dramatic statistical mechanical prediction, spontaneous energy concentration as in star formation, by realizing an aggregate of systems with negative specific heat?

They can. The two-mode Bose-Hubbard (BH) model with repulsive interactions has been highly studied as a model system, representing bosonic particles that quantum mechanically tunnel back and forth between two potential wells \cite{Milb97,Smer97}. It has also been realized as such in experiments \cite{Bloc05,Gati07,Myat97,Matt99}.  In the formalism of so-called second quantization, where the canonical operators $\o[1,2]{a}$ remove atoms from wells 1 and 2, and their conjugate operators $\op[1,2]{a}$ replace the atoms correspondingly, the Hamiltonian of a single two-mode BH system reads
\begin{align}\label{H2}
\frac{\o[2]{H}}{\h} &= -\frac{\Omega}{2}\left(\op[1]{a}\o[2]{a} +\op[2]{a}\o[1]{a} \right) + \frac{\varepsilon\Omega}{2
N}\sum_{\z=1}^{2} \oop[\z]{a}\oo[\z]{a}\;.
\end{align}
For repulsively interacting trapped bosons, the constants $\Omega>0$ and $\varepsilon>0$ are determined by the details of the trapping potential and of interparticle scattering, respectively. $N$ is chosen equal to the expectation value of $\op[1]{a}\o[1]{a}+\op[2]{a}\o[2]{a}$, which represents the total number of particles in both wells together, and is conserved by time evolution under $\o[2]{H}$.

\begin{figure}[tb!]
 \begin{center}
 \includegraphics[width=\linewidth]{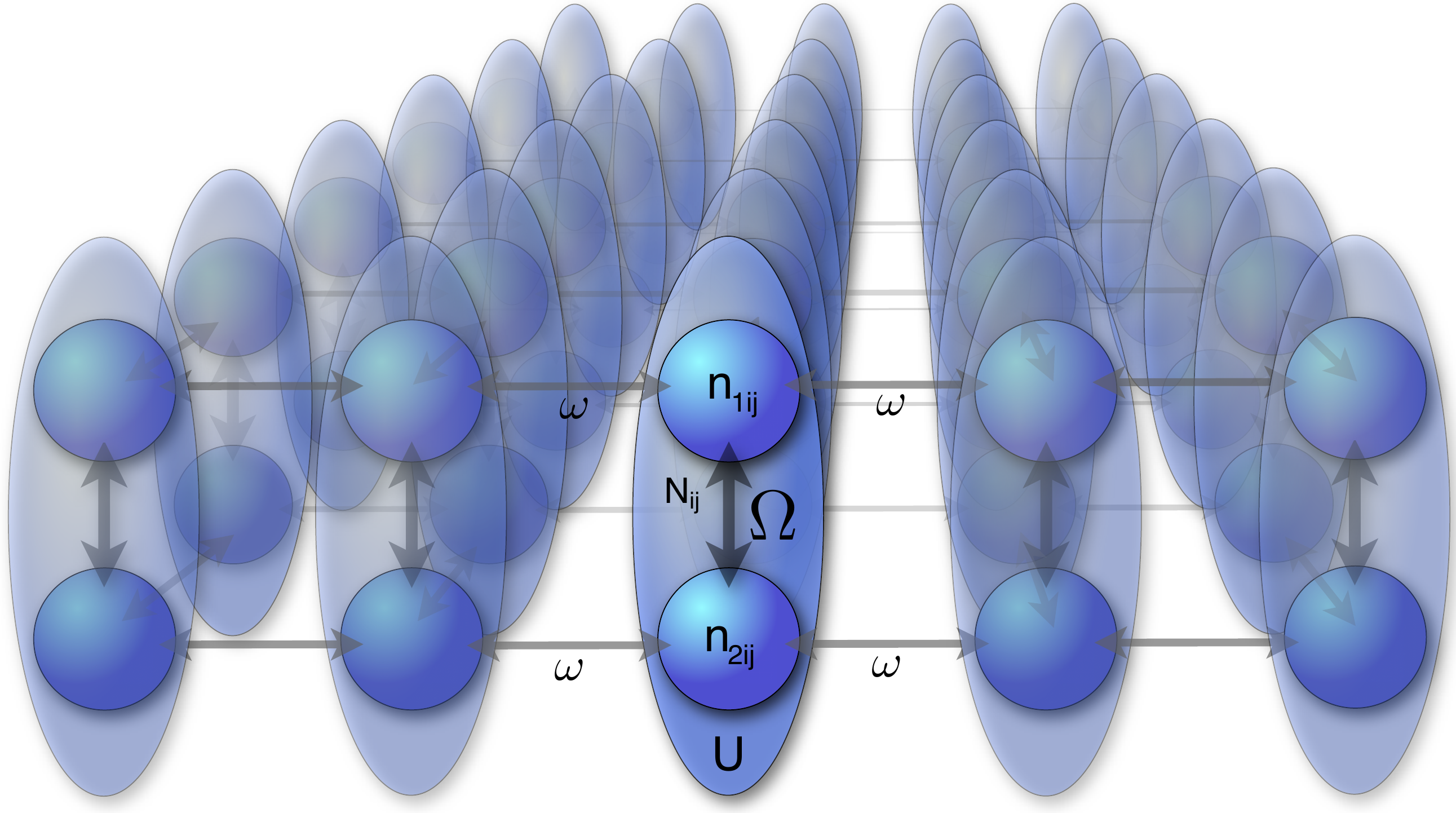} 
 \end{center}
 \caption{(Color online) Two-mode Bose-Hubbard systems weakly coupled in a two-dimensional sheet. Atoms oscillate between modes within each subsystem at the tunnelling Rabi rate $\Omega$, as well as between the subsystems at $\omega \ll \Omega$.}
  \label{systemfig}
\end{figure}

A system with only two degrees of freedom like this is not normally treated statistically, but if we consider a large collection of many such systems, weakly coupled together as sketched in Fig.~\ref{systemfig}, then applying statistical mechanics to the individual two-mode systems is exactly like the standard textbook case of deducing the thermodynamic properties of a dilute gas from the statistical mechanics of a single non-interacting particle. So with precisely the same logic, we obtain the quantum statistical mechanical entropy of two-mode BH in the microcanonical ensemble as
\begin{eqnarray}\label{microcan1}
S_{\Delta E}(E) = k_{B}\ln [Z_{\Delta E}(E) / Z_{\Delta E}(0)]
\end{eqnarray}
where $Z_{\Delta E}(E)$ is the number of eigenstates of $\o[2]{H}$ with eigenvalues in the range $(E,E+\Delta E)$, and $k_{B}$ is Boltzmann's constant. Here the ground state energy and entropy have both been set to zero by shifting all energies and entropies uniformly, as one is free to do in thermodynamics. In the thermodynamic limit of a dense energy spectrum, where this definition is normally applied, $\Delta E$ must be chosen great enough for $Z_{\Delta E}(E)$ to be large and relatively smooth as a function of $E$, but small enough for $Z_{\Delta E}(E)$ to be linearly proportional to $\Delta E$ for all $E$, so that $S_{\Delta E}(E)\to S(E)$ becomes independent of $\Delta E$. 

The two-mode BH spectrum becomes dense in the limit of large total boson number $N$, which is also the regime most easily attained in experiments, and in this semiclassical limit the eigenspectrum is given accurately, except for classical orbits too near an unstable fixed point, by Bohr-Sommerfeld quantization. This involves constructing canonical action-angle variables $J,\phi$ for the quantum system's classical analogue, expressing the classical Hamiltonian $H_{2}$ as $E(J)$, and then determining the quantum eigenenergies as $E_{n} = E(2\pi\h n)$ for $n$ a whole number. The limit of a dense spectrum is obtained for large $N$, and implies $Z_{\Delta E}(E)= (\d J/\d E)\Delta E/(2\pi \h)$. For the classical analogue of $\o[2]{H}$ the derivative $\d J/\d E$ may be calculated analytically in closed form (see the Supplementary Material), yielding the entropy
\begin{align}
S(E) = \kb \ln\left[\sqrt{1+\varepsilon}\frac{2}{\pi}\frac{K\left[k(E/E_{\mathrm{max}},\varepsilon)\right]}{\sqrt{\varepsilon R(E/E_{\mathrm{max}},\varepsilon)}}\right]
\end{align}
where $K(k)$ is the complete elliptic integral of the first kind, and we define the functions
\begin{align}
R(\mathcal{E},\varepsilon) &= \frac{1}{\varepsilon}\sqrt{(1+\varepsilon)^2-4\varepsilon\mathcal{E}},\\
k(\mathcal{E},\varepsilon) &= \frac{1}{2}\sqrt{[\varepsilon-\varepsilon(R(\mathcal{E},\varepsilon) -1/\varepsilon)^2]/R(\mathcal{E},\varepsilon) }\;.
\end{align}
Here $E_{\rm max}(N,\Omega,\varepsilon)$ denotes the highest energy of the classical system for fixed $N$, so that $\mathcal{E}=E/E_{\rm max}$ is the normalized energy. The entropy turns out only to depend on $N$ and $\Omega$ by depending on $\mathcal{E}$. (In this sense the two-mode BH entropy is non-extensive with particle number $N$, but this should not be surprising, since what is expected is extensivity in the number of two-mode systems that are coupled together in a large aggregate system.) The specific heat may then be calculated microcanonically in terms of derivatives of $S(E)$: $C_{V}^{\rm \mu C}=-(\partial S/\partial E)^2/(\partial^2 S/\partial E^2)$.

\begin{figure}[tb!]
\begin{center}
\includegraphics[height=0.49\linewidth]{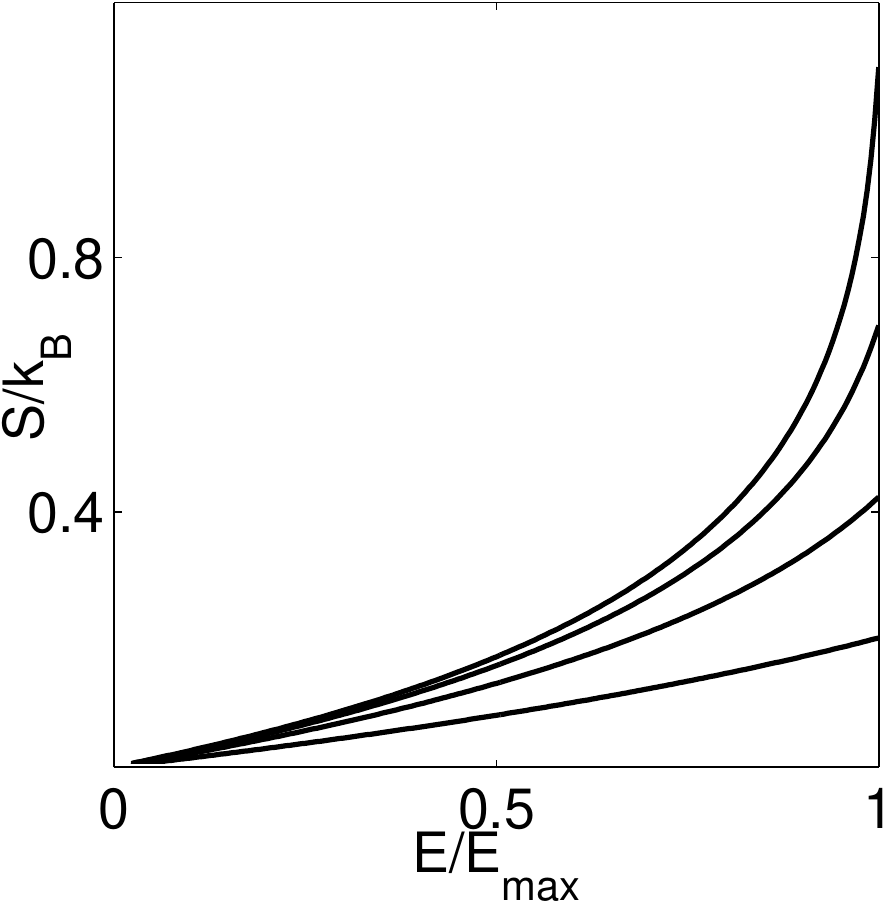}\hspace*{0.1cm} 
\includegraphics[height=0.49\linewidth]{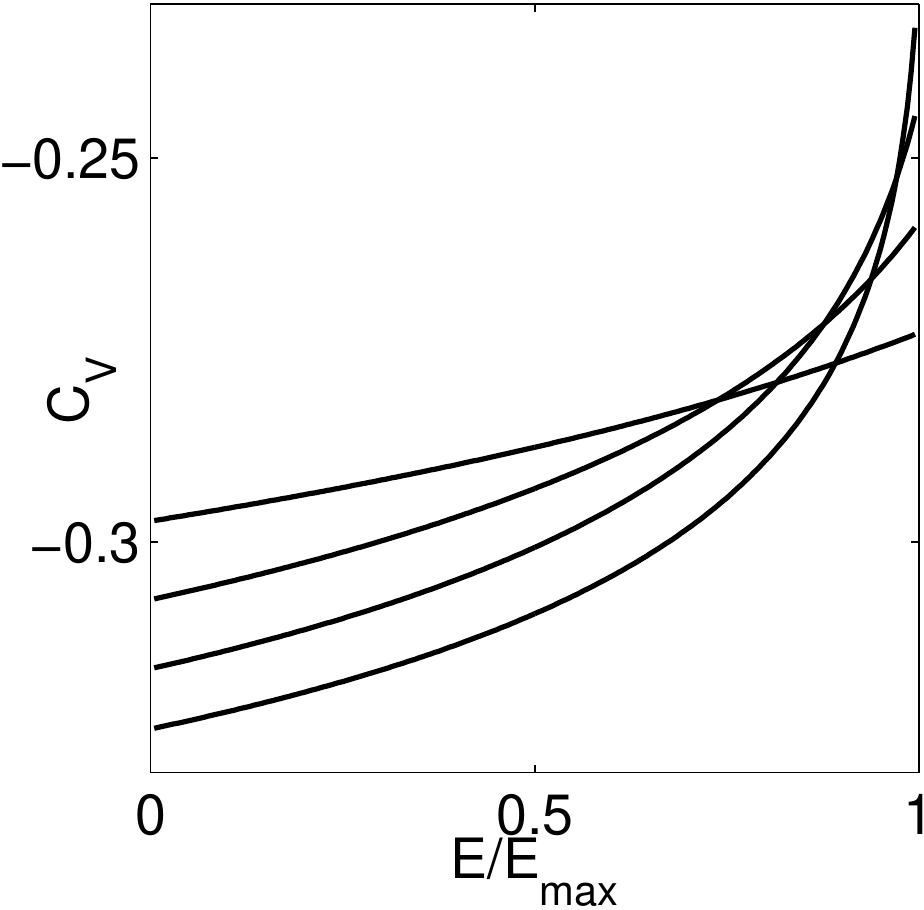}
\caption{Left: Microcanonical Entropy $S(E)$ of the two-mode BH system in units of $\kb$ as a function of normalized energy for $\varepsilon = 0.2, 0.4, 0.6$ and $0.8$ (bottom to top), where ground state energy and entropy are both set to zero. Right: Microcanonical heat capacity $C_V(E)$ of a single two-mode system as a function of energy for the same values of $\varepsilon$ (top to bottom at the left of the graph)}
\label{entropyCV} 
\end{center}
\end{figure}

Both $S$ and $C_{V}^{\rm \mu C}$ are plotted versus $\mathcal{E}$ for different values of $\varepsilon$ in Fig.~\ref{entropyCV}. Note that $C_{V}^{\rm \mu C}$ is \textit{always} negative. For $\varepsilon = 1$, $C_{V}^{\rm \mu C}$ diverges at $\mathcal{E}=1$, but at this point Bohr-Sommerfeld quantization breaks down for any $N$ because the maximum energy classical orbit is a dynamically unstable fixed point. For $\varepsilon < 1$ and large $N$, the statistical mechanical result is definite: an array of weakly coupled two-mode BH systems, as in Fig.~\ref{systemfig}, is an aggregate of interacting subsystems that all have negative $C_{V}$.

Realizing a large array of two-mode BH subsystems, each weakly coupled to its neighbours, is certainly possible in current laboratories that trap cold bosons in so-called optical lattice potentials, induced by standing waves of laser light \cite{Bloc05}. The weak coupling, which allows the subsystems to exchange both atoms and energy with their neighbours, is provided by quantum tunnelling, just as between the two wells within each two-mode subsystem, except that a slower frequency $\omega\ll\Omega$ is imposed, by making the potential barriers higher. 

It is also possible to excite some of these two-mode subsystems locally. If the assumptions of statistical mechanics are correct, then thermodynamics implies that such initially distributed energy will concentrate, as it does in star-producing nebulae, rather than dispersing as it does in almost all other cases, because the weakly coupled subsystems all have negative specific heat. This is a strict test of statistical mechanics, but it is a fair test. Predicting the behaviour of complex aggregate systems like the array shown in Fig.~\ref{systemfig}, from thermodynamic properties of the isolated components of the aggregate as plotted in Fig.~\ref{entropyCV}, is precisely what statistical mechanics is supposed to do. Does it pass the test?

\begin{figure}[tb!]
\begin{center}
\includegraphics[width=0.495\linewidth]{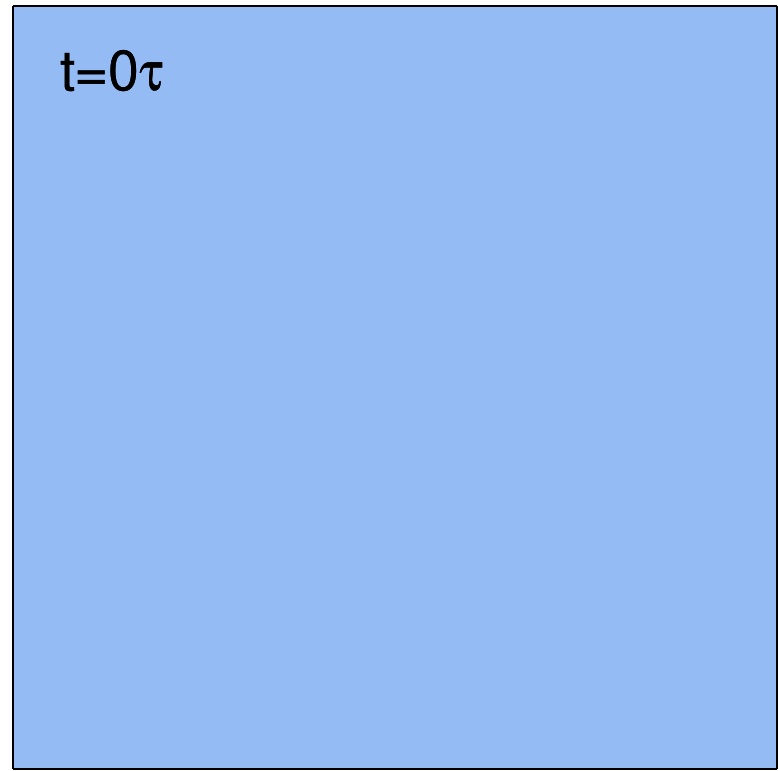}\hspace*{0.1cm} 
\includegraphics[width=0.495\linewidth]{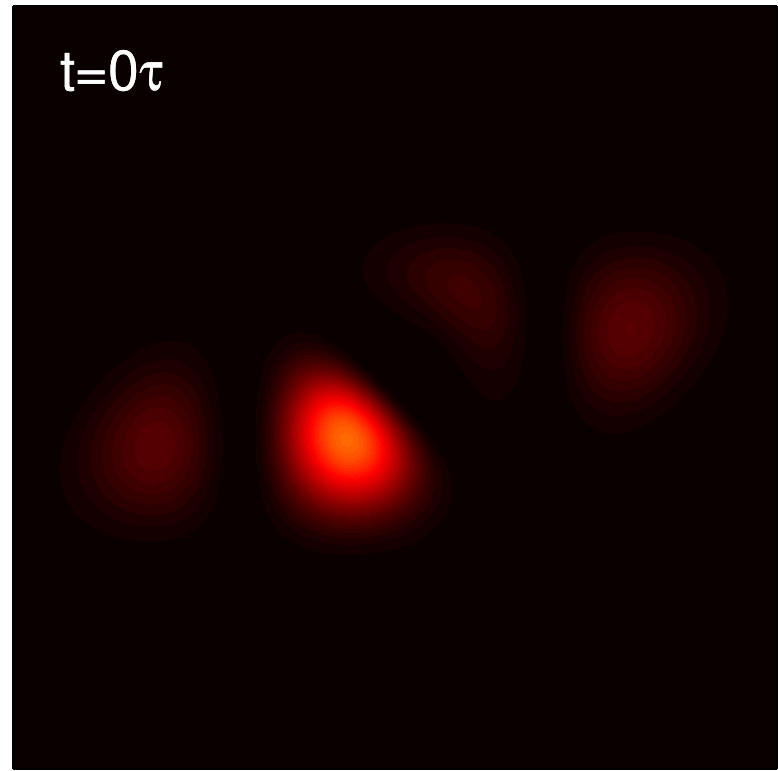}\\\vspace*{0.1cm}
\includegraphics[width=0.495\linewidth]{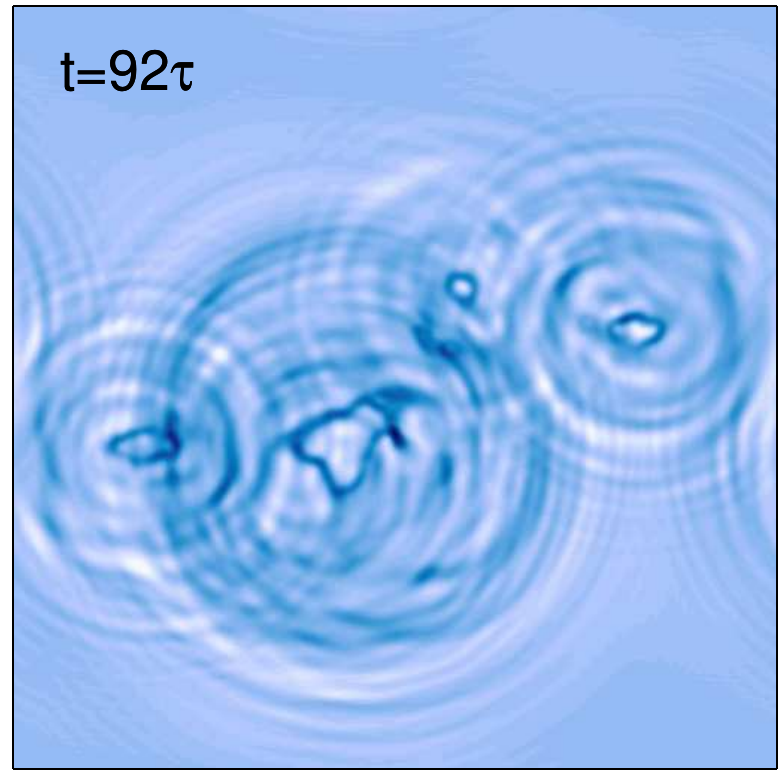}\hspace*{0.1cm}  
\includegraphics[width=0.495\linewidth]{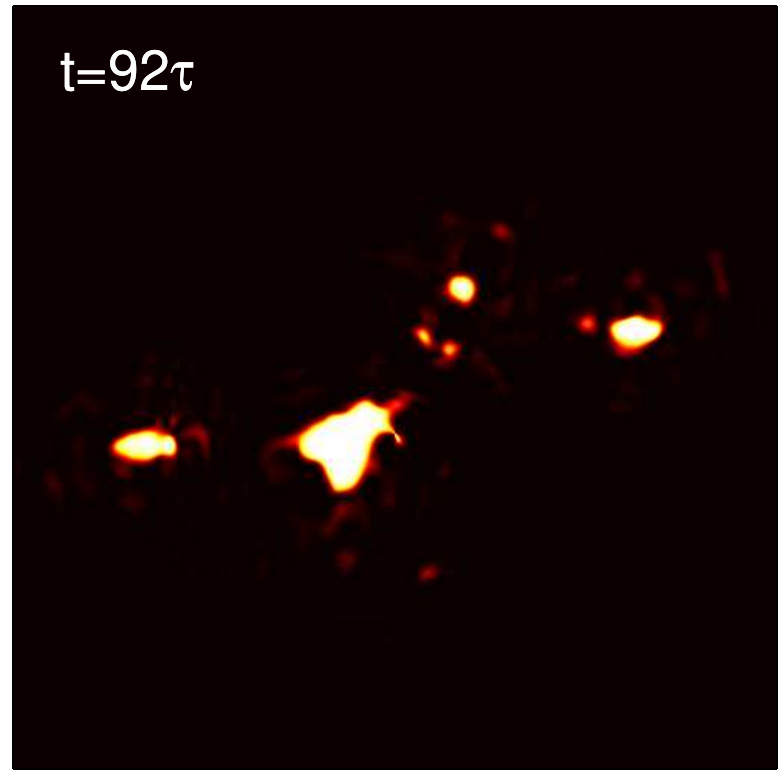}\\\vspace*{0.1cm} 
\includegraphics[width=0.495\linewidth]{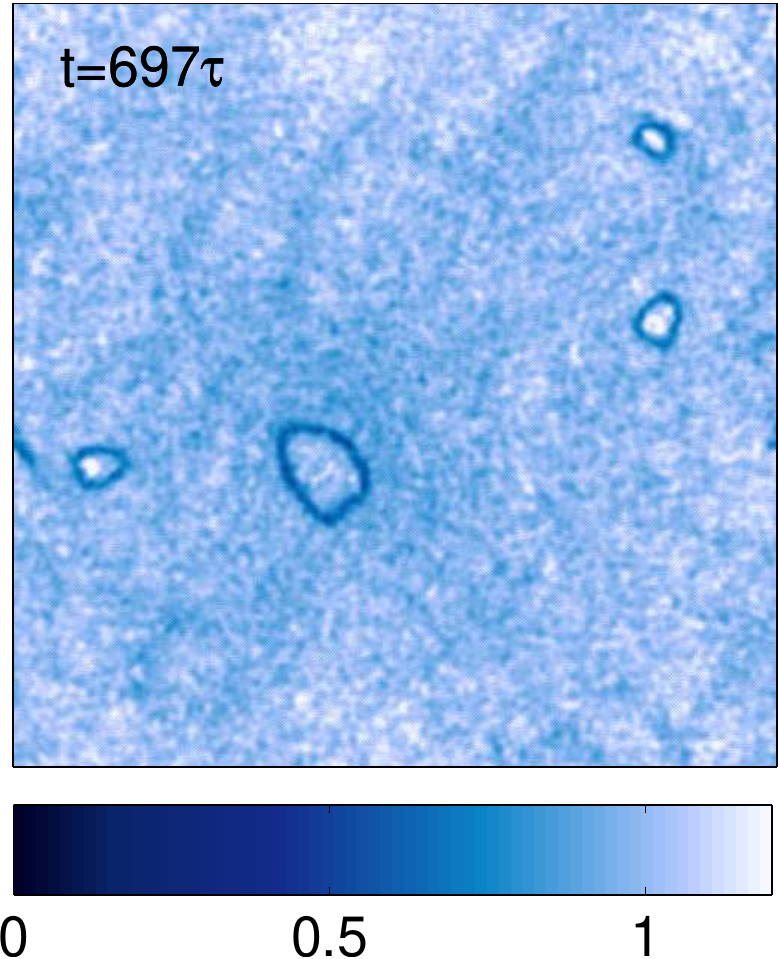}\hspace*{0.1cm}  
\includegraphics[width=0.495\linewidth]{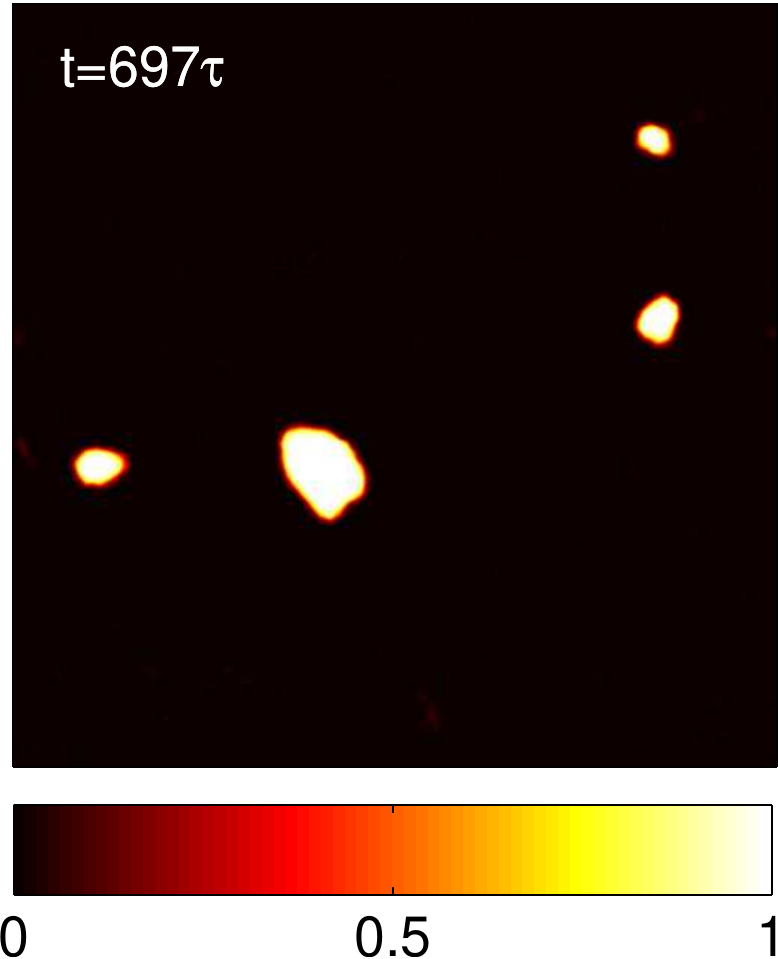} 
\caption{(Colour online) Left column: Total particle number in each two-mode system $N_{ij}$ at three points in time ($t =0\tau, 92\tau$ and $697\tau$, where $\tau = 2\pi/\omega$). Right column: Normalized energy of the two-mode system $E/E_{\rm max}$ at each site at the same points in time. System parameters: $\h\Omega = 1$, $\h\omega = 0.1$, $U = 2$ and $N=1024$, lattice size: $512\times 512$. We start with uniformly distributed atoms and diffuse clouds of excitation defined by simulated phase imprinting. These clouds concentrate into maximally excited droplets.}
\label{numerics}
\end{center}
\end{figure}

In a basic regime we can confirm that it does. The Hamiltonian for the large aggregate system reads
\begin{align}\label{Hamiltonian}
 \o{H} &= \sum_{\x,\y} \o[\x\y]{H} + \o[\x\y]{T},\\
\o[\x\y]{H} &=  -\frac{\h\Omega}{2}\left(\op[1,\x\y]{a}\o[2,\x\y]{a} +\op[2,\x\y]{a}\o[1,\x\y]{a} \right) + U\sum_{\z=1}^{2} \oop[\z,\x\y]{a}\oo[\z,\x\y]{a},\nonumber\\
\o[\x\y]{T}  &=-\frac{\h\omega}{2}\sum_\z\left(\op[\z,\x\y]{a}\o[\z, \x +1,\y]{a} + \op[\z,\x\y]{a}\o[\z,\x, \y +1]{a} + \textrm{H. c.}\right).\nonumber
\end{align}
in which we recognize $\o[ij]{H}$ as a two-dimensional array of $\o[2]{H}$ forms, with the nonlinearity constant written with $U$ rather than $\varepsilon\h\Omega/(2N)$ because $N \to N_{ij}$ can now vary in time, as the $\o[ij]{T}$ term lets particles tunnel between neighbouring subsystems. In the experimentally attainable limit where all the $N_{ij}$ remain large, the complex quantum many-body dynamics of $\o{H}$ can be well approximated with classical mean-field theory \cite{Pita03}. Even the classical dynamics is chaotic, but it can be integrated numerically. The results for a 512-by-512 array of two-mode BH systems are shown in Fig.~\ref{numerics}, as three `stills' from a \href{http://www.physik.uni-kl.de/uploads/media/Strzys-Droplet-movie.mp4}{video} available in the Supplementary Material.

The non-equilibrium initial state, which includes only weak energetic inhomogeneity, relaxes by excitation of sound waves, \Ie waves in the number distribution $N_{ij}$, seen in the left column of Fig.~\ref{numerics}. Some of the initial energy thus disperses into these complex low-amplitude waves; but the majority of the two-mode energy concentrates into bright droplets of maximum excitation, seen in the right column of the Figure. The droplet boundaries are also visible in the left column as domain-wall-like depressions in the local particle number, which extend over several lattice sites. Their thickness is on the order of the characteristic `healing' length of the mean-field theory \cite{Pita03}. The regions of energy concentration behave very much like droplets, with positive surface tension. They can also move (see the Supplementary \href{http://www.physik.uni-kl.de/uploads/media/Strzys-Droplet-movie.mp4}{video}).

The results shown in Fig.~\ref{numerics} and the Supplementary \href{http://www.physik.uni-kl.de/uploads/media/Strzys-Droplet-movie.mp4}{video} are robust and typical over a wide range of parameters and initial conditions. The numerical analysis therefore confirms the dramatic statistical mechanical prediction of spontaneous energy concentration in the semiclassical limit. This is our first main result. Moreover, we are able to follow the formation of `droplets of heat' as a dynamical process. At least in the regime we have analysed here, we can thereby `reverse engineer' the thermodynamics of energy concentration, in the sense that we can identify the dynamical mechanisms underlying the phenomenon.

The weak coupling limit $\omega \ll \Omega$ implies a time scale separation, such that when two neighbouring subsystems exchange energy, the sum of their two action variables $J$ is a so-called adiabatic invariant. In a Bohr-Sommerfeld semiclassical sense, this implies that quanta of local two-mode BH energy are approximately conserved. A resummed Bogoliubov transformation \cite{Strz10,Strz12} shows the same result perturbatively, rather than semi-classically. One can then derive a low-frequency effective theory for the evolution of atoms and plasmon-like two-mode excitations, as two species of separately conserved quasi-particles. This comes near to reviving the 18th century caloric theory of heat \cite{Lavo77}, in a quantum mechanical re-interpretation.

Expanding the full Hamiltonian $\o{H}$ in terms of these new quasi-particles, we find that the leading interactions among the plasmon-like energy quasi-particles are \textit{attractive} whenever those among the atoms are repulsive \cite{Strz10,Strz12}. Within this purely mechanical effective theory, the `heat droplets' predicted by statistical mechanics thus appear as structures like the spin domains seen in spinor condensates \cite{Sadl06}, which are understood in terms of purely mechanical instabilities. In repulsive BH systems, energy concentrates spontaneously because local energy excitations attract each other. 

This promising insight into the microphysical basis of an important thermodynamic phenomenon has only been achieved in a simple limit, however. Beyond the mean-field or perturbative regimes, the quantum many-body theory of large BH arrays with large particle number and high excitation is extremely challenging, even with the best available computations. Spontaneous energy concentration is a robust enough mean-field effect, however, that it must surely extend in some form into more strongly quantum mechanical regimes. Our second main contribution here is to have shown how much can be learned about the mesoscopic interplay between quantum mechanics and thermodynamics, by exploring the dramatic phenomenon of spontaneous energy concentration, with quantum gas experiments on coupled Bose-Hubbard systems. We can bring the heat that kindles stars into the ultracold laboratory.

\section*{Supplementary Material}

\subsection*{Time evolution video}

The Supplementary video of the evolution of the local particle number and the local energy of a $512 \times 512$ lattice of two-mode BH systems can be found online at

\url{http://www.physik.uni-kl.de/uploads/media/Strzys-Droplet-movie.mp4}.

For more information cf.~the caption of Fig.~3.

\subsection{Specific heat in the CE}

The expectation value of the energy in thermal equilibrium according to the CE is given by
\begin{align}
\langle E\rangle_{\rm CE} = \sum_i{E_i \exp{\left(-\beta E_i\right)}}\left[\sum_i{\exp{\left(-\beta E_i\right)}}\right]^{-1}
\end{align}
where $\beta = 1/(\kb T)$. The specific heat may then immediately be calculated according to
\begin{align}
C_V^{\rm CE} &= \frac{\d}{\d T}\langle E\rangle_{\rm CE} = -\kb \beta^2 \frac{\d}{\d \beta}\langle E\rangle_{\rm CE}\nonumber\\
 &= \kb \beta^2 \langle (E - \langle E\rangle)^2\rangle_{\rm CE}
\end{align}
which is equivalent to equation \eqref{CVcanon}.

\subsection{Entropy and negative specific heat}

It is easy to show that two systems with negative heat capacity cannot be in ordinary thermal equilibrium. To see this suppose that total entropy $S$ of two subsystems with total energy $E$ is given by 
\begin{align}
S = S_1(E_1) + S_2(E-E_1)
\end{align}
In equilibrium we should have maximum entropy for given total energy, and we require $\partial S/\partial E_1=0$ to have an entropy extremum, and hence the temperatures of the subsystems  $T_i = (\partial S/\partial E_i)^{-1}$ must be equal, $T_1=T_2$. But this is only equilibrium if this extremum of the entropy is a maximum. Therefore we must consider also the curvature of the entropy, given by
\begin{align}
\frac{\partial^2 S}{\partial E_1^2} = -\left(\frac{1}{T_1^2 C_1}+\frac{1}{T_2^2 C_2}\right)
\end{align}
with $1/C_i = -T_i^2 (\partial^2S/\partial E_i^2)$. If both $C_1<0$ and $C_2<0$, we always have $\partial^2 S/\partial E_1^2>0$, and $S$ has a \textit{minimum} at $T_{1}=T_{2}$. Any entropy-increasing but energy-conserving fluctuations will spontaneously bring the system away from the equal temperature state, by heating up one of the subsystems at the price of cooling the other one down. If the double system does settle down to a maximum entropy state, the two subsystems will be strongly correlated, with one having most of the energy, and the other very little. Such states cannot be described by the CE, because its defining feature is that it makes the subsystems uncorrelated. In an aggregate of many subsystems, however, a set of subsystems which happen all to have nearly equal energies may admit a CE description locally, allowing a canonical definition of temperature that would correspond to the local measurements furnished by a sufficiently small thermometer.

\subsection{Entropy of a two-mode BH system}

To compute the micro-canonical entropy of two-mode BH in the Bohr-Sommerfeld limit, we must construct the classical action-angle variables for this system. It is more convenient to do this by abandoning the explicit creation and destruction operators in favour of angular momentum according to the Schwinger representation:
\begin{align}
\o[x]{L} &= \frac{\h}{2}\left(\op[1]{a}\o[2]{a} +\op[2]{a}\o[1]{a} \right),\\
\o[y]{L} &= \frac{\h}{2\I}\left(\op[1]{a}\o[2]{a} -\op[2]{a}\o[1]{a} \right),\nonumber\\
\o[z]{L} &= \frac{\h}{2}\left(\op[1]{a}\o[1]{a} -\op[2]{a}\o[2]{a} \right).\nonumber
\end{align}
With these definitions \eqref{H2} can be recast into
\begin{align}
\o[2]{H} = -\Omega \o[x]{L} + \frac{\varepsilon\Omega}{N\h}\oo[z]{L} + \frac{\varepsilon\h\Omega}{4N}\o{N}(\o{N}-2), 
\end{align}
where the last term, depending only on $\o{N}$, commutes with the rest of the Hamiltonian. Since it therefore only shifts all energies by a constant, it will be omitted without loss of generality in the following. 

The classical energy of the two-mode BH system thus is equal to
\begin{align}
E &=   -\Omega L_x + \frac{\varepsilon\Omega}{N\hbar} L_z^2.
\end{align}
We then introduce new variables $R\geq 0$ and $\gamma$ through the ansatz
\begin{align}
L_x &= \frac{N\hbar}{2}\left(R\cos(\gamma) - \frac{1}{\varepsilon} \right),\\
L_y &= \frac{N\hbar}{2} R\sin(\gamma),\nonumber\\
L_{z} &=\pm\sqrt{\left(\frac{N\hbar}{2}\right)^{2}-L_{x}^{2}-L_{y}^{2}}\;.
\end{align}
The $\pm$ in the definition of $L_{z}$ means that we use two patches of the $R,\gamma$ co-ordinates to cover the $L_{x,y,z}$ sphere. Note that the requirement that all three of $L_{x,y,z}$ remain real constrains the ranges of $R$ and $\gamma$. In particular, for $\varepsilon<1$, $R$ must lie between $\varepsilon^{-1}- 1$ and $\epsilon^{-1}+1$.

In terms of the new variables the energy reads
\begin{align}\label{2modeE}
E = N\h \Omega\frac{\varepsilon}{4}\left[1+\frac{1}{\varepsilon^{2}} -R^2\right] \equiv E_{\rm max} \mathcal{E}(R,\varepsilon)+E_0,
\end{align}
independent of $\gamma$. This shows that $R$ is a constant of the motion, and so $R,\gamma$ are already a step towards the action-angle variables, and only need to be rescaled.
In (\ref{2modeE}), $E_\mathrm{max}+E_{0}$ and $E_{0}$ are by definition the maximum and minimum values of $E(R,\varepsilon)$, so that $\mathcal{E}(R,\varepsilon)\in [0,1]$ by construction. Since the maximum and minimum of $E$ are attained, respectively, for the minimum and maximum values of $R$, namely $\varepsilon^{-1}\mp 1$, we we see that the maximum excitation energy is $E_{\rm max}=N\hbar\Omega$, and $E_{0}=-N\hbar\Omega/2$. Solving \eqref{2modeE} then for $R(\mathcal{E},\varepsilon)$, we find
\begin{align}
R(\mathcal{E},\varepsilon) &= \frac{1}{\varepsilon}\sqrt{(1+\varepsilon)^2-4\varepsilon\mathcal{E}}.
\end{align}

To determine the proper rescaling that makes true action-angle variables, we now use the canonical equations of motion for the $R,\gamma$ variables, which read
\begin{align}
\dot{R} =0,\quad \dot{\gamma} = \varepsilon\Omega\sqrt{1-\frac{1}{\varepsilon^2}-R^2+\frac{2}{\varepsilon}R\cos(\gamma)}.
\end{align}
Defining the elliptic parameter 
\begin{align}\label{kkdef}
k(\mathcal{E},\varepsilon) &= \frac{1}{2}\sqrt{\frac{\varepsilon}{R}}\sqrt{1-(R-\varepsilon^{-1})^2}.
\end{align}
the equation of motion of the angular variable $\gamma$ can be integrated to yield an analytic expression for the period
\begin{align}
\mathcal{T}(\mathcal{E},\varepsilon) &= \frac{8k(\mathcal{E},\varepsilon) K(k(\mathcal{E},\varepsilon))}{\varepsilon\Omega\sqrt{1-[R(\mathcal{E},\varepsilon)-\varepsilon^{-1}]^2}} = \frac{4 K(k)}{\sqrt{\varepsilon R}\,\Omega}
\end{align}
of a classical orbit; here $K(k)$ is the complete elliptic integral of the first kind. On the other hand, in terms of the properly rescaled action-angle variables $J,\phi$ of the classical system, the canonical equations must read
\begin{align}
\dot{J} &= 0,\quad \dot{\phi} = \frac{\partial E}{\partial J} = \frac{2\pi}{\mathcal{T}(R,\varepsilon)}\;,
\end{align}
This implies that for our system we must have
\begin{align}
\frac{\partial J}{\partial E} =\frac{\mathcal{T}}{2\pi}= \frac{2 K(k)}{\pi\sqrt{\varepsilon R(\mathcal{E},\varepsilon)}\,\Omega}\;.
\end{align}

Using this result we may therefore analytically calculate the number states within $\Delta E$ of $E$ in Bohr-Sommerfeld quantization to be
\begin{align}
Z_{\Delta E}(E) = \frac{\partial J}{\partial E}\frac{\Delta E}{2\pi \h} = \frac{K(k)}{\pi^2\sqrt{\varepsilon R}\,\Omega}\frac{\Delta E}{\h}\;.
\end{align}
Using the identity $K(0)=\pi/2$, we compute the micro-canonical entropy with both entropy and energy of the ground state set to zero to be
\begin{align}
S(E) &= \kb \ln{[Z_{\Delta E}(E)/Z_{\Delta E}(0)]}\nonumber\\
& = \kb \ln\left[\frac{2K(k)}{\pi}\frac{\sqrt{1+\varepsilon}}{\sqrt{\varepsilon R}}\right]
\end{align}
as quoted in our main text, for $k(\mathcal{E},\varepsilon)$ given by (\ref{kkdef}).


\end{document}